\documentclass[epjST]{svjour}
\usepackage{graphics}
\usepackage{amsmath}
\usepackage{amssymb}
\usepackage{epsfig}
\usepackage{bm}
\usepackage{color,wasysym}
\begin{document}
\title{
Critical elasticity at zero and finite temperature
}
\author{Mario Zacharias, 
Achim Rosch, \and Markus Garst\thanks{\email{garst@thp.uni-koeln.de}}}
\institute{Institut f\"ur Theoretische Physik, Universit\"at zu K\"oln,\\
Z\"ulpicher Str. 77, 50937 K\"oln, Germany}
\abstract{
Elastic phase transitions of crystals and phase transitions whose order parameter couples linearly to elastic degrees of freedom are reviewed with particular focus on instabilities at zero temperature. A characteristic feature of these transitions is the suppression of critical fluctuations by long-range shear forces. As a consequence, at an elastic crystal symmetry-breaking quantum phase transition the phonon velocity vanishes only along certain crystallographic directions giving rise to critical phonon thermodynamics described by a stable Gaussian fixed point. At an isostructural solid-solid quantum critical end point, on the other hand, the complete suppression of critical fluctuations results in true mean-field critical behavior without a diverging correlation length. Whenever an order parameter couples bilinearly to the strain tensor, the critical properties are eventually governed by critical crystal elasticity. This is, for example, the case for quantum critical metamagnetism but also for the classical critical Mott end point at finite $T$. We discuss and compare the solid-solid end points expected close to the Mott transition in  V$_2$O$_3$ and $\kappa$-(BEDT-TTF)$_2 X$.
} 
\maketitle
\section{Introduction}
\label{intro}

In most quantum or classical phase transitions in solids, the underlying lattice is affected by criticality. The change of the lattice constant measured in a thermal expansion experiment is, for example, often a very sensitive probe of criticality. In this review, we will focus on situations where either the lattice itself becomes critical or strongly affects criticality. 
The ratio of thermal expansion, $\alpha = (1/V)(\partial V/\partial T)_p$, and specific heat, $C_p = T (\partial S/\partial T)_p$, turns out to be a useful, unbiased tool to identify and classify zero temperature quantum phase transitions in general, that can be tuned with pressure. It can be identified with the relative change of temperature upon adiabatically varying the pressure
\begin{eqnarray}
\Gamma = \frac{1}{V_m T} \left.\frac{\partial T}{\partial p}\right|_S=\frac{\alpha}{C_p},
\end{eqnarray}
with the molar volume $V_m$, and it is closely related to the parameter originally introduced by Gr\"uneisen \cite{Grueneisen:1908}. The Gr\"uneisen ratio $\Gamma$ is predicted to diverge with a characteristic power-law at a quantum critical point \cite{Zhu03}. At the critical pressure $p_c$, it diverges as a function of temperature as $\Gamma \sim T^{-1/(\nu z)}$ with an exponent determined by the product of the correlation length exponent $\nu$ and the dynamical exponent $z$. Moreover, in the low-temperature limit the divergence as a function of pressure is universal, $\Gamma = - \frac{G}{V_m(p-p_c)}$, in the sense that its prefactor $G$ is solely determined by critical exponents. The Gr\"uneisen parameter thus not only provides a necessary criterion for the existence of a quantum critical point but also yields important information about its universality class by identifying certain combinations of critical exponents. Moreover, $\Gamma$ also changes sign close to a quantum critical point, and the sign changes reveal the locations of entropy acummulation \cite{Garst:2005}.
Measurements of $\Gamma$ \cite{Kuechler:2003,Kuechler:2004,Kuechler:2006,Lorenz:2007} and its magnetic counterpart, the magnetocaloric effect \cite{Wolf:2011,Tokiwa:2012,Tokiwa:2013a,Tokiwa:2013b,Tokiwa:2014,Ryll:2014}, have proven useful in the investigation of various quantum critical systems, see Refs.~\cite{Loehneysen:2007,Gegenwart:2008,Gegenwart:2010} for reviews.

However, the theory of Ref.~\cite{Zhu03} implicitly assumes that the crystal lattice couples only perturbatively to the critical system and can be considered as a non-invasive probe that does not modify the critical properties itself. This assumption is often justified.  
Nevertheless, there are exceptions where the crystal lattice and critical degrees of freedom are instead strongly coupled and the assumptions of Ref.~\cite{Zhu03} break down. Such quantum phase transitions that are governed by crystal elasticity are at the focus of this short review. We will argue that the non-perturbative coupling to the crystal lattice has a profound effect and fundamentally changes the critical behavior. The reason for this is that the rigidity of the crystalline lattice can induce long-ranged forces which suppress critical fluctuations, reminiscent of the suppression of Goldstone modes by long-ranged gauge interactions within the Higgs mechanism. Consider, for example, a large fluctuating domain of size $L$ close to a critical point where the lattice constant within this domain is changed by a factor $1+\epsilon$ with $\epsilon \ll 1$. If the product $L \epsilon$ is much larger than a lattice constant $a$, the crystal becomes strongly strained. As a result, the corresponding energy cost can suppress the formation of such domains that exceed a certain size. It can, however, happen that the whole crystal deforms without such large scale fluctuations in a transition described asymptotically by mean-field theory.

For classical phase transitions, the coupling of critical fluctuations to the elastic degrees of freedom, following early work by Rice \cite{Rice:1954} and Domb \cite{Domb:1956}, were extensively studied in the 1970ies \cite{Levanyuk:1970,Villain:1970,Larkin:1969,Sak:1974,Wegner:1974,Bergman:1976,Aubry1971,Cowley:1976,Folk:1976,Folk79,Schwabl:1980,Schwabl:1996,Chalker1980}. Some of the results presented here are straightforward generalizations of these previous works to the case of quantum phase transitions. We start in section \ref{subsec:basics} with a short introduction to elasticity theory and recapitulate in section \ref{subsec:ElastoCouplings} the most important elastic couplings to the critical fluctuations. Section \ref{sec:QCE} reviews Ref.~\cite{Zacharias2014} and discusses quantum critical elasticity focussing  in particular on spontaneous crystal symmetry-breaking quantum phase transitions and isostructural solid-solid quantum critical end points (QCEPs). In section \ref{sec:MQCEP}, we discuss quantum critical metamagnetism and argue that it is generally preempted by critical elasticity providing an example of a fluctuation-induced solid-solid QCEP. Finally, in section \ref{sec:MEDP} the critical properties of the Mott end point at finite temperature is reviewed, which is in fact an example of a classical solid-solid end point. We distinguish between a strong and a weak non-perturbative elastic coupling scenario that might be relevant for V$_2$O$_3$ and $\kappa$-(BEDT-TTF)$_2 X$, respectively. Some of the results for the Mott end point have been already presented in Ref.~\cite{Zacharias12}. Part of this review is also based on Ref.~\cite{ZachariasPhD}.

\subsection{Basics of elasticity theory}
\label{subsec:basics}

The fundamental quantities of elasticity theory are introduced in the following, which are required for the subsequent discussion. For an in-depth introduction we refer the reader to standard textbooks \cite{LandauBook,BornBook}.

The static elastic properties of a crystal are described by the following potential for the strain  tensor field $\varepsilon_{ij}$ 
\begin{align}
\mathcal{V}_{\rm elast} = \frac{1}{2} \varepsilon_{ij} C_{ijkl} \varepsilon_{kl} 
\end{align}
where $C_{ijkl}$ is the elastic constant tensor. It is convenient to decompose the strain field into a macroscopic part $E_{ij}$ and a part that carries finite momentum,
\begin{align} \label{decomposition}
\varepsilon_{ij}({\bf r}) = E_{ij} + \frac{1}{2}\Big(\partial_i u_j({\bf r}) + \partial_j u_i({\bf r}) \Big).
\end{align}
The latter is related to the phonon displacement field $u_i$. The low-energy phonon Lagrangian reads 
\begin{align} \label{Lphonon}
\mathcal{L}_{\rm phonon} = \frac{\rho}{2} (\partial_t u_i)^2 - \frac{1}{2} \partial_i u_j C_{ijkl} \partial_k u_l
\end{align}
with the mass density $\rho$. The phonon energy-momentum dispersion follows from the Lagrangian \eqref{Lphonon} and is given by $\rho \omega^2 - q_i C_{ijkl} q_k = 0$. In particular, the phonon velocities are determined by the eigenvalues of the dynamical matrix $M_{j l}({\bf q}) = q_i C_{ijkl} q_k$. On the other hand, the macroscopic strain is governed by the potential
\begin{align} \label{ElasPot}
\mathcal{V}(E_{ij}) = \frac{1}{2} E_{ij} C_{ijkl} E_{kl} + \sigma_{ij} E_{ij}
\end{align}
where we also allowed for the presence of a macroscopic stress tensor $\sigma_{ij}$. In case of an applied hydrostatic pressure, $p$, the stress tensor is diagonal $\sigma_{ij}= - p \delta_{ij}$. For a uniaxial pressure, say, along the $x$-direction, $\sigma_{ij}= - p_x \delta_{ix}\delta_{jx}$ etc.
The macroscopic strain also possesses kinetic energy, and the boundary conditions couple its equation of motion to the one of the phonon modes. Both effects do not contribute in the thermodynamic limit and will therefore be neglected in the following. 

Generally, the strain $E_{kl}$ can be decomposed into irreducible representations of the crystal class. Depending on the degeneracy of the corresponding eigenvalue of the elastic constant matrix, these representations can be either singlets, doublets or triplets. For example, a cubic crystal possesses a singlet $E_{xx}+E_{yy}+E_{zz}$, a doublet $(E_{xx}-E_{yy}, 2 E_{zz}-E_{xx}-E_{yy})$ and a triplet $(E_{yz}, E_{xz}, E_{xy})$ with eigenvalues of the elastic constant matrix in Voigt notation $C_{11}+2 C_{12}$, $C_{11} - C_{12}$, and $C_{44}$, respectively.
Macroscopic stability of the crystal requires that each of these eigenvalues are positive, i.e., in general the tensor $C_{ijkl}$ should be positive definite so that the potential \eqref{ElasPot} is stable.

\subsection{Elastic coupling to critical fluctuations}
\label{subsec:ElastoCouplings}

We now consider the coupling of these elastic degrees of freedom to the critical fluctuations of a second-order phase transition. In order to be concrete, let us assume that this transition is described by the following $\phi^4$-potential in terms of a real singlet order parameter $\phi$
\begin{align} \label{crPot}
\mathcal{V}_{\rm cr} = \frac{r}{2} \phi^2  + \frac{u}{4!} \phi^4 
- h \phi
\end{align}
where $u>0$ and the critical point obtains for $r=0$ and $h = 0$. The order parameter $\phi$ might represent some magnetic, electronic, or other degrees of freedom. 
The crystal lattice that houses a critical subsystem will unavoidable couple to the fluctuations of the order parameter 
$\phi$. Let us consider the most important couplings that are linear in the strain field. As we consider here only a singlet order parameter $\phi$, this coupling can only involve a certain singlet of the crystal group that we denote by $\varepsilon$, and the interaction can be written in the form
\begin{align} \label{ElastoCoupling}
\mathcal{V}_{\rm int} = \frac{1}{2} \gamma_2 \varepsilon \phi^2 - \gamma_1 \varepsilon \phi.
\end{align}
Whereas the coupling $\gamma_2$ is expected to be finite, the bilinear coupling $\gamma_1$ is often forbidden by symmetry. For example, for an Ising ferromagnet where $\phi$ corresponds to the Ising magnetization the coupling $\gamma_1$ must vanish due to time-reversal symmetry. Similarly, order parameters that break translational symmetries cannot couple linearly to the strain. The coupling $\gamma_1$ is however finite, for example, at end points of 
lines of first-order metamagnetic or Mott transitions, or at an electronic nematic transition.  

\subsubsection{Perturbative treatment of the elastic coupling}
\label{subsec:PertElastoCoupling}

In the perturbative regime, we can neglect the phonon contributions and consider only the macroscopic strain singlet, $E$, in Eq.~\eqref{ElastoCoupling}. For a cubic crystal, for example, the elastic part then reduces to 
\begin{align}
\mathcal{V}_{\rm el} = \frac{1}{2} \gamma_2 E \phi^2 - \gamma_1 E \phi + \frac{K}{2} E^2 - p E
\end{align}
where $K$ is the corresponding modulus, i.e., $K = (C_{11} + 2 C_{12})/3$, and $p$ is the hydrostatic pressure. Perturbatively minimizing for the strain singlet one obtains $E \approx p/K$ and the effective critical potential assumes the same form as Eq.~\eqref{crPot} 
\begin{align}
\mathcal{V}_{\rm eff} = \frac{r(p)}{2} \phi^2 + \frac{u}{4!} \phi^4
- h(p) \phi
\end{align}
but with pressure-dependent parameters, $r(p) = r + \gamma_2 p/K$ and $h(p) = h + \gamma_1 p/K$. The tuning parameters, i.e., the prefactors of the most relevant operators in the critical theory, $h$ and $r$, thus obtain an effective pressure dependence. 

Let us first discuss the case where $\gamma_1$ is forbidden and, furthermore, $h = 0$.
The contribution to the free energy attributed to the quantum critical degrees of freedom can then be written in the scaling form
\begin{align} \label{CriticalF}
\mathcal{F}_{\rm cr} = T^{\frac{d+z}{z}} f\Big(\frac{r(p)}{T^{1/(\nu z)}}
\Big),
\end{align}
where the scaling function $f$ is assumed to possess the following asymptotic behavior $f(x) \to const.$ for $x \to 0$ and $f(x) \sim |x|^{\nu (d+z)}$ for $|x| \to \infty$. This scaling form for the quantum critical free energy with a pressure dependent tuning parameter, $r$, was used in Ref.~\cite{Zhu03} for the discussion of the Gr\"uneisen parameter.
However, the critical free energy \eqref{CriticalF} also 
predicts a critical renormalization of the bulk modulus. At $T=0$ it reduces to $\mathcal{F}_{\rm cr} \propto |r|^{\nu(d+z)}$ and the renormalization of the bulk modulus reads
\begin{align}
\delta K \sim \partial^2_p \mathcal{F}_{\rm cr} \sim \partial^2_r \mathcal{F}_{\rm cr} \sim |r|^{\nu(d+z) - 2}
\end{align}
This renormalization will be perturbative as long as 
\begin{align} \label{Criterion}
\nu(d+z) - 2 > 0.
\end{align}
This is just the generalization of the corresponding result for classical criticality where the elastic coupling remains perturbative if the specific heat exponent is negative \cite{Rice:1954,Domb:1956,Larkin:1969}, i.e., $\alpha  = 2 - \nu d < 0$, from which the criterion for quantum criticality \eqref{Criterion} obtains after replacing the dimension $d$ with the enhanced dimension $d+z$. This criterion simply arises from the requirement that the perturbative renormalizations of the elastic constants due to the correlator $\langle \phi^2 \phi^2 \rangle$ of the order parameter remains non-singular. 
Although there might be exceptions for effectively low-dimensional systems, see e.g.~Ref.~\cite{Anfuso:2008}, the criterion \eqref{Criterion} for quantum criticality is usually fulfilled. In fact, the theory of Ref.~\cite{Zhu03} for the Gr\"uneisen parameter implicitly assumes that the criterion \eqref{Criterion} is obeyed.

\subsubsection{Non-perturbative elastic coupling to critical degrees of freedom}

The elastic coupling $\gamma_2$ becomes non-perturbative if the criterion \eqref{Criterion} is violated. On the other hand, the bilinear coupling $\gamma_1$ will always be non-perturbative. In the latter case, the corresponding perturbative correction to the bulk modulus is attributed to the correlator $\langle \phi \phi \rangle$ of the order parameter that is singular by definition. 
Note that this also applies to situations where the order parameter $\phi$ itself is a bilinear in fermionic fields 
and the critical theory must be formulated in terms of fermionic degrees of freedom. 
Whenever a coupling $\gamma_1$ is present it will lead to a strong entanglement between the critical and elastic degrees of freedom so that a perturbative treatment of the elastic coupling unavoidably breaks down. If no first-order transition is induced, one obtains instead an {\em elastic} critical point.
The properties of such continuous elastic phase transitions is at the focus of the following chapters. In chapter \ref{sec:QCE} we start with a general analysis of continuous elastic quantum phase transitions irrespective of their origin, i.e., whether they are genuine or are induced by other, critical degrees of freedom. The effective critical elastic theory is analysed, the quantum critical thermodynamics, and, in particular, the behavior of the Gr\"uneisen parameter is discussed. In chapter \ref{sec:MQCEP}, the particular example of metamagnetic quantum criticality is investigated that generically allows for a bilinear coupling $\gamma_1$ and is thus non-perturbative. Finally, in chapter \ref{sec:MEDP} a corresponding example of classical criticality is also discussed where we review
the issue of the universality class of the critical end point at finite temperature for the Mott transition.

\section{Quantum critical elasticity}
\label{sec:QCE}

The instabilities of a crystal lattice can be studied from the point of view of an effective critical theory in terms of the strain order parameter. Such an approach was chosen in Ref.~\cite{Zacharias2014} where the quantum critical signatures of continuous elastic quantum phase transitions were discussed building on previous work by Cowley \cite{Cowley:1976} and Schwabl and collaborators \cite{Folk:1976,Folk79,Schwabl:1980,Schwabl:1996}. At such an instability an eigenvalue of the elastic constant matrix $C_{ijkl}$ vanishes. The strain order parameter is identified as the corresponding eigenvector, and depending on the degeneracy of this eigenvalue it can either be a singlet, doublet or a triplet of the irreducible representations of the crystal class. 

Importantly, if such an eigenvalue of $C_{ijkl}$ vanishes the phonon velocities in general remain finite. Mathematicaly this arises because the eigenvalues of $C_{ijkl}$ are determined from the $6 \times 6$ matrix of the elastic constant matrix in Voigt notation, $C_{\rho \lambda}$, whereas the phonon velocities derive from the eigenvalues of the dynamical matrix $M_{j l}({\bf q}) = q_i C_{ijkl} q_k$ (see above). Physically, it reflects the fact that general deformations of the crystal cannot be represented in terms of an acoustic phonon mode for fixed direction, $\hat q = {\bf q}/|{\bf q}|$.
Only for specific directions of the phonon wavevector $\hat q$, the polarization of the phonons might induce the particular type of strain that is akin to the one of the strain order parameter so that the phonon velocity only vanishes for this particular direction $\hat q$ in momentum space. As discussed by Cowley \cite{Cowley:1976} and Folk {\it et al.} \cite{Folk:1976}, there are elastic transitions where the phonon velocities either remain finite for all $\hat q$, vanish for $\hat q$ pointing along certain directions or lying within specific two-dimensional planes of momentum space. The phonons thus become only critical in a $m$-dimensional subspace with $m=0,1,2$ corresponding to type 0, I or II, respectively, in the classification of Cowley \cite{Cowley:1976}. As result of this restricted critical subspace, even the classical critical theory is either at or above its upper critical dimension for $m=2$ and $m=0,1$, respectively \cite{Folk:1976,Chalker1980}. We will see below that the corresponding quantum phase transition is always above its upper critical dimension. In the following, we distinguish between elastic transitions that do and do not break a crystal symmetry. 

\subsection{Spontaneous symmetry-breaking elastic transitions}
\label{subsec:SSBET}

\begin{table}[t]
\centering
\begin{tabular}{|c|c|c|c|}
\hline
elastic transition & constant & strain & type
\\
\hline
orthorhombic $\to$ monoclinic & $c_{44}$ & $\varepsilon_{23}$ & I  
\\
orthorhombic $\to$ monoclinic & $c_{55}$ & $\varepsilon_{13}$ & I  
\\
orthorhombic $\to$ monoclinic & $c_{66}$ & $\varepsilon_{12}$ & I  
\\
tetragonal $\to$ orthorhombic & $c_{11}-c_{12}$ & $\varepsilon_{11}-\varepsilon_{22}$ & I  
\\
tetragonal $\to$ orthorhombic & $c_{66}$ & $\varepsilon_{12}$ & I  
\\
tetragonal $\to$ mono- or triclinic & $c_{44}$ & $(\varepsilon_{23}, \varepsilon_{13})$ & I+II  
\\
hexagonal $\to$  mono- or triclinic & $c_{44}$ & $(\varepsilon_{23}, \varepsilon_{13})$ & I+II   
\\
\hline
\end{tabular}
\caption{Table from Ref.~\cite{Zacharias2014} based on results of Refs.~\cite{Cowley:1976,Folk:1976,Schwabl:1980} listing 
possible second-order elastic phase transitions. The columns from left to right specify the crystal symmetry that is broken at the elastic transition, the eigenvalue of the elastic constant matrix in Voigt notation that vanishes at the transition, the strain order parameter and the type of the transition in the classification of Cowley \cite{Cowley:1976}. Certain modifications apply for tetragonal crystals with a finite component $c_{16}$.
}
\label{table1}
\end{table}

For elastic transitions that break a crystal symmetry the expectation value of the macroscopic strain order parameter $E$ is zero in the symmetric undistorted phase and assumes a finite value in the symmetry-broken phase. Such a transition can be discontinuous or continuous, i.e., first or second order, respectively, depending on the properties of the Landau potential $\mathcal{V}(E)$ for the order parameter. Assuming that $\mathcal{V}(E)$ is analytic so that a Taylor expansion around $E = 0$ can be performed, the transition can be of second order whenever the cubic term in this expansion vanishes. According to Refs.~\cite{Cowley:1976,Folk:1976,Schwabl:1980}, this is the case for the transitions listed in Table~\ref{table1}. With the exceptions listed in the last two rows in Table~\ref{table1}, most of these transitions are characterized by a singlet strain order parameter $E$, for which the potential reads
\begin{align} 
\mathcal{V}(E) = \frac{r}{2} E^2 + \frac{u}{4!} E^4 + \sigma E,
\end{align}
where $\sigma$ is the appropriate singlet of an applied stress, which explicitly breaks the symmetry $E \to -E$, and a cubic term $E^3$ is absent. If the quartic coupling is positive $u>0$, a second-order elastic quantum phase transition occurs at $T=0$ and $\sigma=0$ when the tuning parameter goes to zero $r \to 0^+$. This is the case when the elastic constant listed in the second column of Table~\ref{table1} vanishes. The tuning parameter $r = r(T,p)$ in general depends on temperature, $T$, and hydrostatic pressure, $p$. The latter effectively arises from anharmonicities in the elastic potential that couples the singlet $E$ to the trace of the strain tensor, $\mathcal{V}_{\rm anha} \sim {\rm tr}\{\varepsilon_{ij}\} E^2$. Minimizing the trace in the presence of a hydrostatic pressure results in a pressure dependence of $K$ similarly as discussed in section \ref{subsec:PertElastoCoupling}. The temperature dependence, on the other hand, is induced either by anharmonicities, i.e., phonon excitations or coupling to other degrees of freedom, e.g., electrons.

The symmetry-breaking elastic transition are either of type I or II in the Cowley classification. The wavevector of the phonon can be decomposed, $\bf q = (\bf p, \bf k)$, into an $m$-dimensional soft component $\bf p$ and a $(3-m)$-dimensional stiff component ${\bf k}$ with $m=1,2$ for type I and II, respectively. Close to the elastic phase transition, the phonon dispersion then possesses the anisotropic form \cite{Folk79}
\begin{align} \label{CrDispersion}
\omega^2 \sim r {\bf p}^2 + a {\bf p}^4 + b {\bf k}^2 + \dots
\end{align}
with finite constants $a$ and $b$, and the dots represent further terms that are less relevant. For the identification of scaling exponents, it is convenient to perform the substitution ${\bf k}^2 \to {\bf k}'^4$. This effectively corresponds to the introduction of an enhanced spatial dimensionality $d_{\rm eff} = m +2 (d-m) = 2 d - m$ with $d=3$. The 
dispersion \eqref{CrDispersion} is then characterized by the scaling $r \sim {\bf p}^2$ and $\omega^2 \sim {\bf p}^4, {\bf k}'^4$, which determines the correlation length exponent $\nu = 1/2$ and the dynamical exponent $z=2$, respectively.

For classical elastic transitions, the upper critical dimension of the theory is obtained from the criterion $d_{\rm eff} = 4$ which amounts to $d^+_{\rm CPT} = 2 + m/2$ \cite{Folk:1976,Chalker1980}. 
While classical transitions with $m=0,1$ are above their upper critical dimension $d > d^+_{\rm CFT}$ for $d=3$ allowing for a mean-field description, classical type II transitions with $m=2$ are at their upper critical dimension giving rise to logarithmic corrections to mean-field exponents. For elastic quantum phase transitions, on the other hand, the upper critical dimension is further reduced by the dynamical exponent, $d^+_{\rm QPT} = 2 + m/2 - z = m/2$. All elastic second-order quantum phase transition are therefore above their upper critical dimension for $d=3$ and the phonon excitations can be treated perturbatively.

\begin{figure}
\centering 
\begin{minipage}{.4\columnwidth}
\includegraphics[width=\columnwidth]{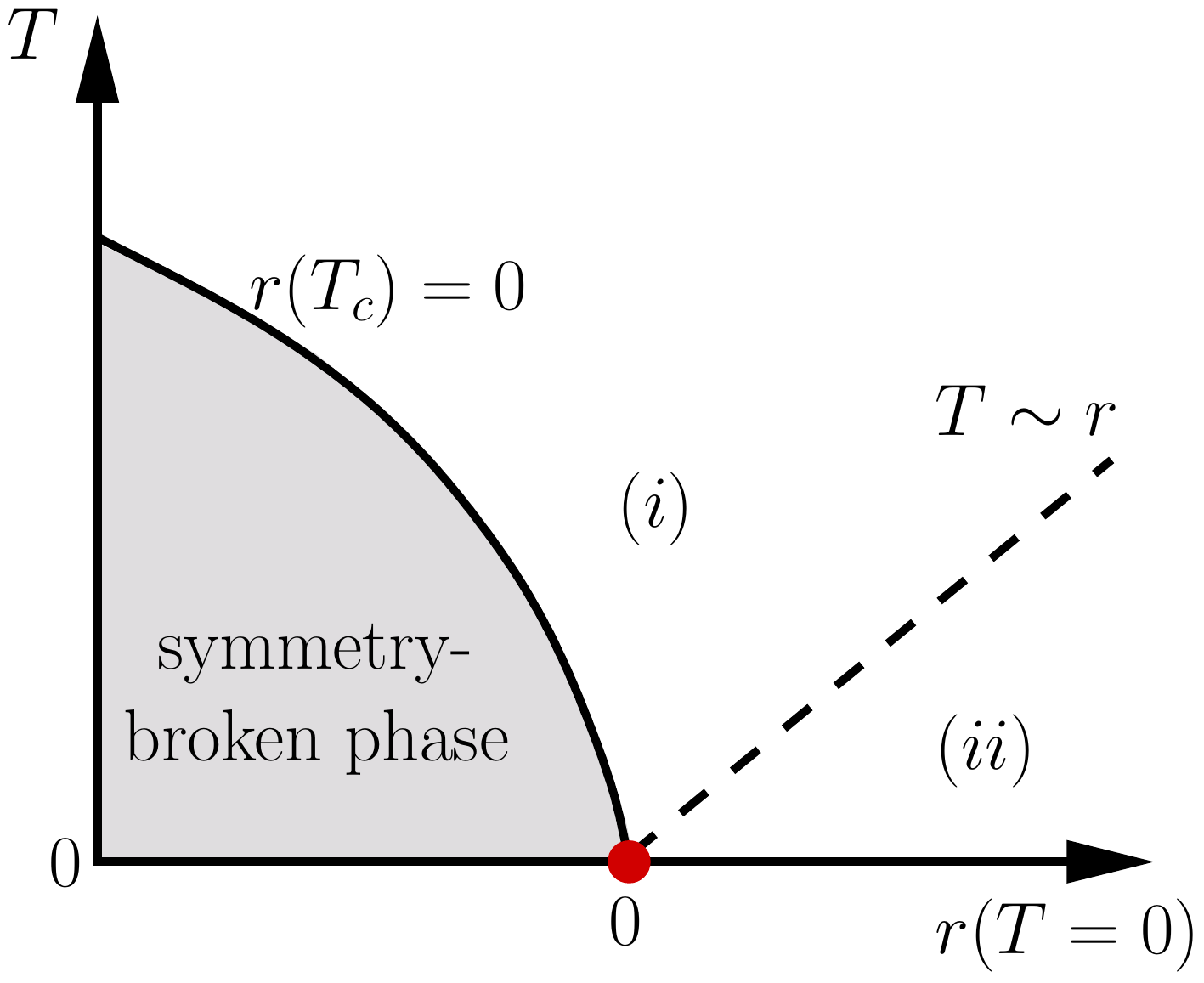}
\end{minipage}
\hspace{4em}
%
%
\begin{minipage}{.4\columnwidth}
\begin{tabular}[c]{|c|c|c|}
\hline
I & $(i)$ & $(ii)$
\\
\hline
$C_{\rm cr} \sim$ & $T^{5/2}$ & $T^3 r^{-1/2}$
\\
$\alpha_{\rm cr} \sim$ & $T^{3/2}$ &  $T^3 r^{-3/2}$
\\
$\alpha_{\rm cr}/C_{\rm cr}$ & $\sim 1/T$ & $= \frac{1}{6} \frac{1}{p-p_c}$
\\
\hline
\end{tabular}\\[0.3cm]
\begin{tabular}[c]{|c|c|c|}
\hline
II & $(i)$ & $(ii)$
\\
\hline
$C_{\rm cr} \sim$ & $T^{2}$ & $T^3 r^{-1}$
\\
$\alpha_{\rm cr} \sim$ & $T$ &  $T^3 r^{-2}$
\\
$\alpha_{\rm cr}/C_{\rm cr}$ & $\sim 1/T$ & $= \frac{1}{3} \frac{1}{p-p_c}$
\\
\hline
\end{tabular}
\end{minipage}
\caption{ 
{\em Left:} Phase diagram for a symmetry-breaking elastic quantum phase transition. The parameter $r(T=0)$ can be tuned by pressure and vanishes for tuning the corresponding elastic constant to zero, see second column of Table \ref{table1}. Its temperature dependence $r(T)$ determines the shape of the finite-temperature phase boundary. The thermodynamics attributed to the critical phonons shows a crossover at $T \sim r$ resulting in two regimes $(i)$ and $(ii)$. 
{\em Right:} Critical phonon thermodynamics: specific heat $C_{\rm cr}$, thermal expansion $\alpha_{\rm cr}$ and Gr\"uneisen ratio in the regimes $(i)$ and $(ii)$ for pressure tuning $r(T=0) \propto p-p_c$; for type I (upper table) and II (lower table), respectively.
} 
  \label{fig:PhaseDiagram1}
\end{figure}

In particular, the contribution of the critical phonon excitations to the free energy can be written in the scaling form \cite{Zhu03}
\begin{align} \label{Fcr}
\mathcal{F}_{\rm cr} = T^{\frac{d_{\rm eff} +z}{z}} f\Big(\frac{r}{T^{1/(\nu z)}}\Big),
\end{align}
with $\nu z =1$ and $\frac{d_{\rm eff} +z}{z} = 4 - m/2$. An explicit calculation shows that the function $f$ possesses the asymptotics $f(x) = const.$ for $x \to 0$ and $f(x) \sim x^{\nu d_{\rm eff} - \nu z d} = x^{-m/2}$ for $x\to \infty$. From Eq.~\eqref{Fcr} the critical phonon thermodynamics is easily derived. At criticality $r=0$, i.e., in regime $(i)$ of Fig.~\ref{fig:PhaseDiagram1}, the critical contribution to the phonon specific heat is given by 
$C_{\rm cr} \sim T^{3-m/2}$, i.e., $C_{\rm cr} \sim T^{5/2}$ and $C_{\rm cr} \sim T^{2}$ for type I and type II transitions, respectively. The volume thermal expansion, $\alpha$, derives from the pressure dependence of the tuning parameter $r$ so that $\alpha_{\rm cr} \sim T^{2-m/2}$ at $r=0$. For the critical Gr\"uneisen ratio defined as $\Gamma_{\rm cr} = \alpha_{\rm cr}/C_{\rm cr}$ follows \begin{equation} \Gamma_{\rm cr}  \sim \frac{1}{T^{1/(\nu z)}}=\frac{1}{T}\end{equation} with $\nu z = 1$ in agreement with scaling considerations \cite{Zhu03}. In the low-temperature regime $(ii)$ of Fig.~\ref{fig:PhaseDiagram1}, the divergence is universal \begin{equation}\Gamma_{\rm cr} = \frac{m}{6} \frac{1}{V_m(p-p_c)}\end{equation} with the prefactor $m/6$, i.e., $1/6$ and $1/3$ for type I and II, respectively, 
where we used $r(T=0,p) \propto p-p_c$ with the critical pressure $p_c$.  
The critical phonon thermodynamics is summarized in the Tables of Fig.~\ref{fig:PhaseDiagram1}. 

It should be noted, however, that the critical phonon thermodynamics  vanishes with a relatively high power of temperature for $T \to 0$. As a consequence, it might be actually subleading compared to other non-critical contributions, e.g., due to gapless electronic degrees of freedom in metals. In such a case, the identification of the critical phonon parts, $C_{\rm cr}$ and $\alpha_{\rm cr}$, and thus the analysis of the critical Gr\"uneisen ratio $\Gamma_{\rm cr}$ might be challenging.

\subsection{Isostructural elastic transitions}
\label{subsec:IsostructuralElTr}

\begin{figure}
\centering
\resizebox{0.3\columnwidth}{!}{
 \includegraphics{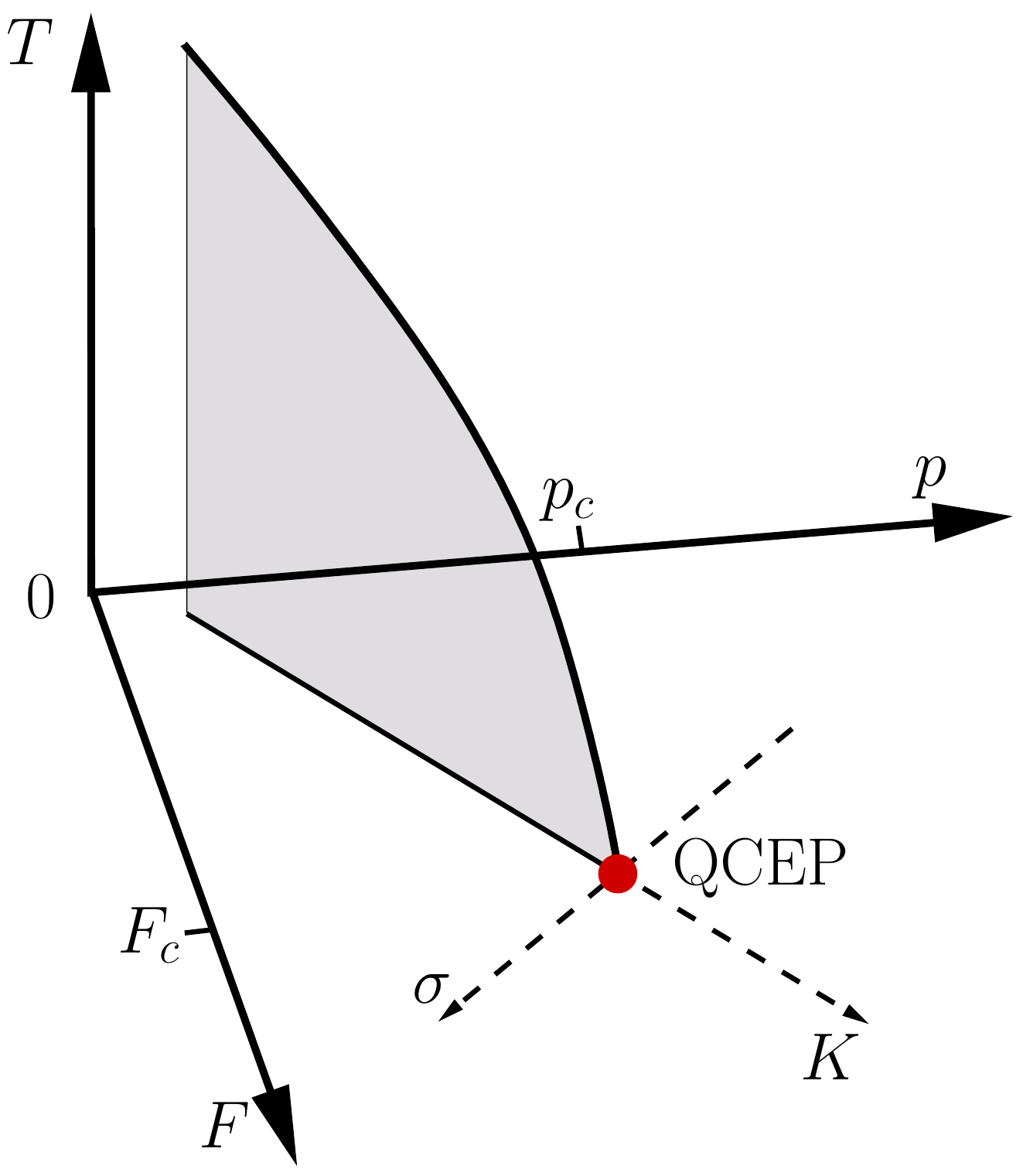}}
\caption{
Phase diagram with a solid-solid quantum critical end point (QCEP). A line of first-order solid-solid quantum phase transitions meets a line of finite temperature second-order transitions at the QCEP and confine a surface (shaded area) of first-order transitions at finite $T$. The dashed coordinate system is defined by the parameters $K$ and $\sigma$, see Eq.~\eqref{SS-Potential}, that vanish at the QCEP.
}
\label{fig:SS-QCEP}       
\end{figure}

Elastic transitions not listed in Table~\ref{table1} are generically first-order and exceptions require additional fine-tuning. For example, the Taylor expansion for the potential of the strain order parameter might allow for a cubic term whose prefactor can, however, be tuned to zero. Another example, on which we focus in the following, are isostructural transitions. The corresponding macroscopic strain order parameter, $E$, is a singlet representation that is invariant under all crystal symmetry operations. Its Landau potential generally includes all powers of $E$; after shifting the order parameter by a constant $E \to E + E_0$ one can eliminate the cubic term and the potential assumes the form 
\begin{align} \label{SS-Potential}
\mathcal{V}(E) = \frac{K}{2} E^2 + \frac{K_4}{4!} E^4 + \sigma E,
\end{align}
with $K_4>0$. For $K < 0$ a first-order phase transition arises where the expectation value of $E$ jumps at $\sigma = 0$ as function of $\sigma$. This corresponds to a transition between two different solids that possess the same crystal symmetries while they are distinguished by different values of $E$. A second-order {\it solid-solid quantum critical end point} (QCEP) obtains when  both  parameters, $K$ and $\sigma$, are tuned to zero at $T=0$. Here, the expectation value of $E$ changes in a critical manner, for example, as a function of $\sigma$. The tuning might be achieved with the help of pressure $p$ and an external field $F$, on which the parameters depend, $K = K(F,p)$ and $\sigma=\sigma(F,p)$; the position of the QCEP within the $(F,p)$ phase diagram is then identified by $K(F_c,p_c) = \sigma(F_c,p_c) = 0$, see Fig.~\ref{fig:SS-QCEP}.

Minimizing the potential \eqref{SS-Potential} one obtains 
\begin{eqnarray}
E \sim \left\{\begin{array}{ll}
\sigma/K & \  \text{for } \ |\sigma| \ll \sqrt{K^3/K_4} \\
 - |\sigma/K_4|^{1/3} {\rm sgn}(\sigma) & \ \text{for } \  |\sigma| \gg \sqrt{K^3/K_4}
 \end{array}\right.
 \end{eqnarray} 
In the latter limit at $F=F_c$, the strain 
depends on the pressure  $E \sim (p-p_c)^{1/3}$ in a non-linear fashion with mean-field exponent $\delta=3$. This corresponds to a breakdown of Hooke's law, which is a hallmark of solid-solid end points. 

Importantly, the solid-solid QCEP is characterized by the absence of critical microscopic fluctuations as the velocity of sound remains finite for all directions in this case. In fact, all isostructural transitions are of type 0 in the Cowley classification and the phonon sector remains non-critical because of the high symmetry of the order parameter $E$. Consequently, the quantum phase transition associated with the solid-solid QCEP is a true mean-field transition and exactly described by Landau theory. 
In particular, there does not exist a diverging correlation length and the usual scaling hypothesis for the critical free energy does not apply. The critical thermodynamics is in fact non-universal and depends on other, non-critical degrees of freedom that induce an effective temperature dependence of the tuning parameters, $K = K(F,p,T)$ and $\sigma=\sigma(F,p,T)$. For example, assuming that $\sigma(F,p,T) = \sigma_0(F,p) + a T^x$, with, e.g., $x=2$ in a metal, it follows from minimizing the potential \eqref{SS-Potential} a critical contribution to the specific heat $C_{\rm cr} \sim |\sigma_0|^{1/3} T^{x-1}$ and thermal expansion $\alpha_{\rm cr} \sim |\sigma_0|^{-2/3} T^{x-1}$ for $T\to 0$ at $K=0$. In the same limit one obtains for the critical Gr\"uneisen ratio $\Gamma_{\rm cr} = \alpha_{\rm cr}/C_{\rm cr}$, 
\begin{align} \label{GrueneisenSS-point}
\Gamma_{\rm cr}\Big|_{T=0, F=F_c} 
= \frac{1}{3(x-1)} \frac{1}{V_m(p-p_c)}
\end{align}
that diverges with the inverse of $\sigma_0(F_{\rm c},p) \propto p-p_c$ in agreement with expectations but with a prefactor depending on the exponent $x$.

As the solid-solid QCEP separates two solids with different volumes but with the same crystal symmetry, the discussion given above is in fact relevant for all quantum critical end points in solids that are not associated with the breaking of a true symmetry. Such critical end points are described with the help of a scalar order parameter $\phi$ and are in the Ising universality class. However, in a crystal this order parameter will in general couple linearly to the strain singlet, i.e., $\mathcal{V}_{\rm int} = - \gamma_1 \phi  \varepsilon$, so that the diverging susceptibility $\langle \phi \phi \rangle$ will cause a isostructural instability with the concomitant change in the universality class.
As will be shown below, in this case the instability of the macroscopic strain $E$ preempts the critical endpoint of the $\phi$ field in such a way that  critical fluctuations are absent at the transition.
This applies, for example, to the Kondo volume collapse transition at zero temperature, where the asymptotic critical behavior will be governed by a solid-solid QCEP \cite{Dzero2006/10,Hackl2008}. 
It is also the case for the metamagnetic QCEP \cite{Millis02,Zacharias13}, where $F$ corresponds to a magnetic field, which will be discussed in some detail in the next chapter. Similar reasoning applies to a metaelectric QCEP, where the Ising order parameter is identified with the variation in the longitudinal electric polarization, $\phi = P - P_{\rm cr}$, which increases in a critical manner at some finite electric field. 
For the corresponding classical metaelectric end point, that exists at finite temperatures in KH$_{2}$PO$_{4}$, the absence of a diverging correlation length has been already demonstrated in Ref.~\cite{Courtens1973}. Another famous classical analogue is the $\gamma-\alpha$ transition of Ce \cite{Poniatovskii1958,Lawrence1975,Croft1977}. In chapter \ref{sec:MEDP}, we will discuss the Mott end point at finite $T$ that is also predicted to be governed by critical elasticity \cite{Zacharias12}.

\section{Metamagnetic quantum critical end point}
\label{sec:MQCEP}

Metamagnetism is associated with a superlinear rise of the magnetization $M(H)$ at some finite magnetic field $H_m$ \cite{Wohlfarth62}. As a function of decreasing temperature such a metamagnetic crossover might develop into a first-order transition where the magnetization instead jumps as a function of field. The resulting line of metamagnetic first-order transitions terminates in a critical end point $(T_{\rm ep},H_m)$ located at finite temperature $T_{\rm ep}$. If this end point temperature $T_{\rm ep}$ can be suppressed by some additional tuning, a metamagnetic QCEP arises for $T_{\rm ep} \to 0$.

Metamagnetic quantum criticality was introduced in the context of Sr$_3$Ru$_2$O$_7$ in order to account for the anomalous behavior close to its metamagnetic field \cite{Grigera01,Millis02,Gegenwart06}. The universal critical thermodynamics close to such a metamagnetic QCEP has been discussed  by Weickert {\it et al.} in Ref.~\cite{Weickert10} 
however without considering the feedback of the lattice to the critical behavior. 
Moreover, Zacharias and Garst \cite{Zacharias13} analyzed in detail the critical thermodynamics close to such a QCEP within the spin-fluctuations theory as proposed in Ref.~\cite{Millis02}. In particular, it was pointed out that the metamagnetic QCEP is intrinsically unstable with respect to a magnetoelastic coupling. The metamagnetic Ising order parameter is given by the change of the longitudinal magnetization close to the critical field $\phi = M - \langle M \rangle_{H=H_m}$. Its Ising symmetry is only emergent and, as a consequence, a linear coupling $\gamma_1$ to the strain is generically allowed. This will have the consequence that the metamagnetic QCEP will be preempted by an isostructural quantum phase transition.

\subsection{Solid-solid quantum critical end point induced by metamagnetic fluctuations}
\label{sec:solid-solidMetamag}

The effective bosonic theory for quantum critical metamagnetism in a metal reads \cite{Millis02}
\begin{align} \label{TheoryMetamagnetism}
\mathcal{L} = \frac{1}{2} \phi \left[r - \nabla^2 \right] \phi + \frac{u}{4!} \phi^4 - h \phi + \mathcal{L}_{\rm dyn}
\end{align}
where $u>0$, $h \propto H-H_m$ measures the distance to the critical field, and the metamagnetic quantum critical end point obtains for $r=0$.
The dynamical part
\begin{align}
\mathcal{S}_{\rm dyn} = \int_0^\beta d\tau \int d \vec r\, \mathcal{L}_{\rm dyn} = \frac{1}{\beta}\sum_{\omega_n {\bf k}}\frac{|\omega_n|}{|\bf k|} |\phi({\bf k}, i\omega_n)|^2 
\end{align}
is governed by the Landau-damping, i.e., damping of the order parameter by particle-hole pairs in the metal. 
The resulting thermodynamics was discussed in detail in Refs.~\cite{Zacharias13,Gegenwart06}. 

A linear coupling $\gamma_1$ of the order parameter to the strain field, see Eq.~\eqref{ElastoCoupling}, is generally allowed and will modify the quantum critical properties. An exhaustive discussion of the theory \eqref{TheoryMetamagnetism}
linearly coupled to the crystal lattice including an analysis of phonon degrees of freedom can be found in Ref.~\cite{ZachariasPhD}. As the resulting isostructural transition will be of type 0 where the phonons remain non-critical, see section \ref{subsec:IsostructuralElTr}, we can limit ourselves here to a coupling of the order parameter to the macroscopic singlet, $E$, of the strain tensor only. Effectively, the coupling $\gamma_1$ can then be absorbed into a shift of the tuning parameter $h \to h + \gamma_1 E$. Integrating out the order parameter fluctuations one then obtains the effective potential for the macroscopic strain $E$,

\begin{align}
\mathcal{V}_{\rm eff}(E) = \frac{K_0}{2}E^2  
+ \mathcal{F}_{\rm cr}(r,h+\gamma_1 E,T)
\end{align}
where $K_0$ is the corresponding modulus and $\mathcal{F}_{\rm cr}(r,h,T)$ is the critical free energy of Eq.~\eqref{TheoryMetamagnetism}. The magnetoelastic coupling gives rise to a renormalization of the modulus 
of the form 
\begin{align}
K \equiv \partial_E^2 \mathcal{V}_{\rm eff}(E)\big|_{E=0}=  K_0 - \gamma_1^2 \chi_{\rm cr} 
\end{align}
where $\chi_{\rm cr} = \langle \phi \phi\rangle = -\partial^2_h \mathcal{F}_{\rm cr}$ is the singular part of the differential susceptibility. At $h=0$ and zero temperature this susceptibility is simply given by $\chi_{\rm cr} = 1/r$ for $r>0$. Consequently, before the metamagnetic QCEP is reached for $r=0$ the renormalized bulk modulus vanishes at $r^* = \gamma_1^2/K > 0$. The metamagnetic critical point is effectively preempted by an isostructural QCEP that is realized for $r^* = \gamma_1^2/K_0$.

Using the results for $\mathcal{F}_{\rm cr}(r,h,T)$ of Ref.~\cite{Zacharias13}, one finds for the effective potential at $r=r^*$
the explicit form close to the isostructural QCEP
\begin{align} \label{PotentialMetamagnet}
\mathcal{V}_{\rm QCEP}(E) 
&= \frac{K_0}{2}E^2  + \frac{K_4}{4!} E^4 + \sigma_0 E - \frac{(h+\gamma_1 E)^2}{2 R^*} + f_{0}(r^*,T)
\end{align}
where we introduced the corresponding singlet stress $\sigma_0$ that couples to $E$ and also explicitly added a quartic term with coefficient $K_4$ that stabilizes the potential (that is also renormalized by $\delta K_4 \sim u \gamma_1^4/{R^*}^4$). The coefficient $R^*$ parameterizes the effective stiffness, $\chi_{\rm cr} = 1/R^*$, of the metamagnetic potential at $r=r^*$,
\begin{align} 
R^*
&=  r^* + u
\left\{
\begin{array}{ll}
\mathfrak{r}_1\, T^{\frac{d+1}{3}}, 
& 
\quad R^* \ll T^{2/3}
\\
\mathfrak{r}_2\, T^2 {r^*}^\frac{d-5}{2}, & 
\quad
R^* \gg T^{2/3}.
\end{array}
\right.
\end{align}
The remaining free energy independent of $E$ is given by
\begin{align} 
f_{0}(r^*,T) = 
\left\{
\begin{array}{cc}
\displaystyle
-\mathfrak{f}_5\, T^{\frac{d}{3}+1},  
& \quad r^* \ll T^{2/3}
\\[1em]
\displaystyle
-\mathfrak{f}_6\, T^2 {r^*}^{\frac{d-3}{2}},
& \quad r^* \gg T^{2/3}.
\end{array}
\right.
\end{align}
The coefficients $\mathfrak{f}_i$ and $\mathfrak{r}_i$ are explicitly listed in Ref.~\cite{Zacharias13}.
The potential \eqref{PotentialMetamagnet} has the same form as the one given in Eq.~\eqref{SS-Potential} with parameters $K = K_0 - \gamma_1^2/R^*$ and $\sigma = \sigma_0 - \gamma_1 h /R^*$. 

A hallmark of the isostructural QCEP is the non-linear response of the crystal lattice with respect to an externally applied stress $\sigma_0$ corresponding to the breakdown of Hooke's law. Minimization of the potential at $T=0$ and $h=0$ yields $E \propto \sigma_0^{1/3}$. On the other hand, the metamagnetic fluctuations that drive the isostructural transition are reflected in a crossover at a characteristic temperature $T \sim (r^*)^{3/2}$. The specific heat for example, that follows from $f_{\rm cr}(r^*,T)$ in Eq.~\eqref{PotentialMetamagnet}, shows an anomalous $T$-dependence, $C \sim T^{\frac{d}{3}}$ at high temperatures, $T \gg (r^*)^{3/2}$, whereas it is of Fermi-liquid type, $C \sim T {r^*}^{\frac{d-3}{2}}$, for $T \ll (r^*)^{3/2}$.
Apart from the contribution to thermodynamics deriving from the underlying metamagnetic fluctuations, $f_{0}(r^*,T)$, there will be also additional $T$-dependences of the bare parameters. In particular, the bare stress $\sigma_0$ will possess an intrinsic $T^2$ temperature dependence in a metal, $\sigma_0(T) = \sigma_0(0) + a T^2$, resulting in a thermal expansion of Fermi-liquid type, $\partial_T E = \partial_T \sigma_0/K_0 \sim T$,  in the absence of criticality. Close to the solid-solid QCEP, however, this will result in an additional, critical contribution to the specific heat. For example, at $h=0$ and at lowest temperatures, $C_{\rm cr} \sim  T \partial^2_T |\sigma_0(T)|^{4/3} \sim |\sigma_0(0)|^{1/3} T$. Similarly, for the thermal expansion we obtain in the same limit $\alpha_{\rm cr} \sim |\sigma_0(0)|^{-2/3} T$ reproducing the result for the critical Gr\"uneisen parameter of Eq.~\eqref{GrueneisenSS-point} with $x=2$. Note, however, that the critical contributions are here subleading compared to the ones deriving from $f_{\rm 0}(r^*,T)$. In order to obtain the expected scaling for the critical Gr\"uneisen parameter $\Gamma_{\rm cr} = \alpha_{\rm cr}/C_{\rm cr}$ a careful subtraction of the leading non-critical contributions are necessary. This limits the practicability of the Gr\"uneisen analysis in the present case.

\subsection{Deviation of metamagnetic correlations from Ornstein-Zernicke form}

\begin{figure}
\centering
(a) 
\resizebox{0.35\columnwidth}{!}{
\includegraphics{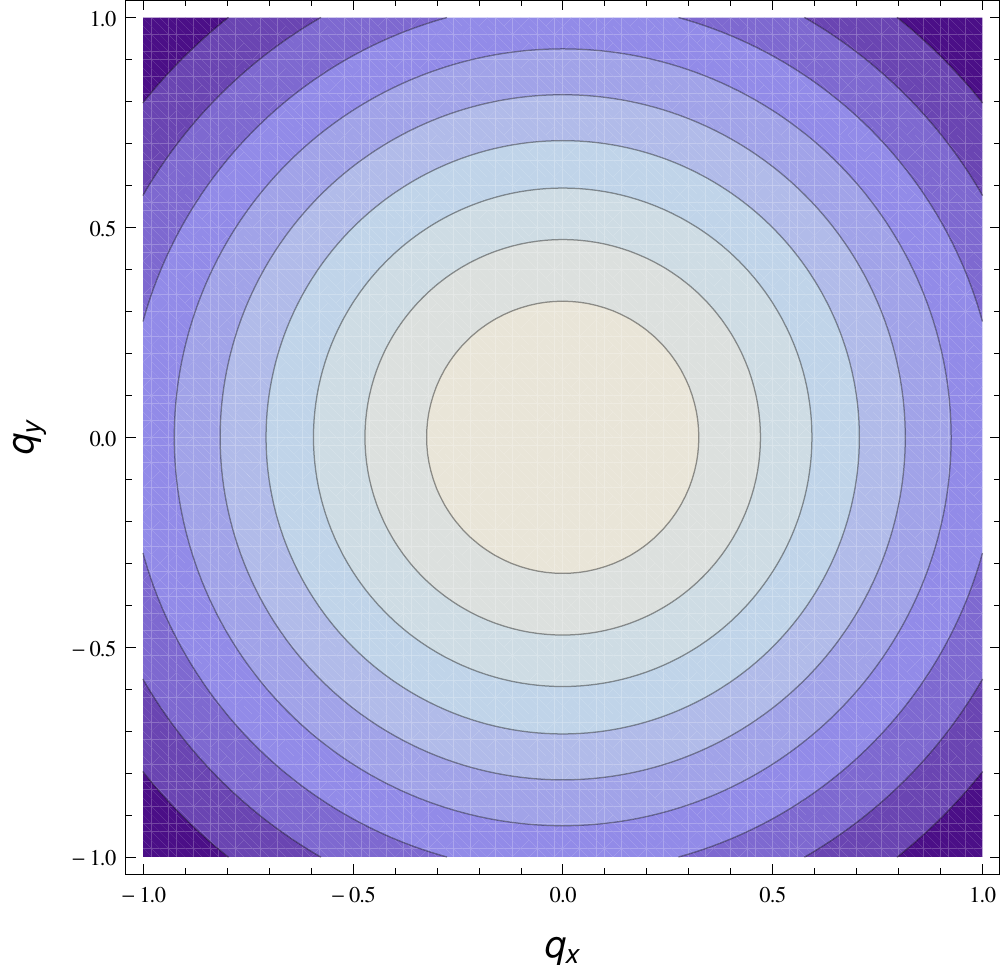}}
\qquad\qquad (b) 
\resizebox{0.35\columnwidth}{!}{
\includegraphics{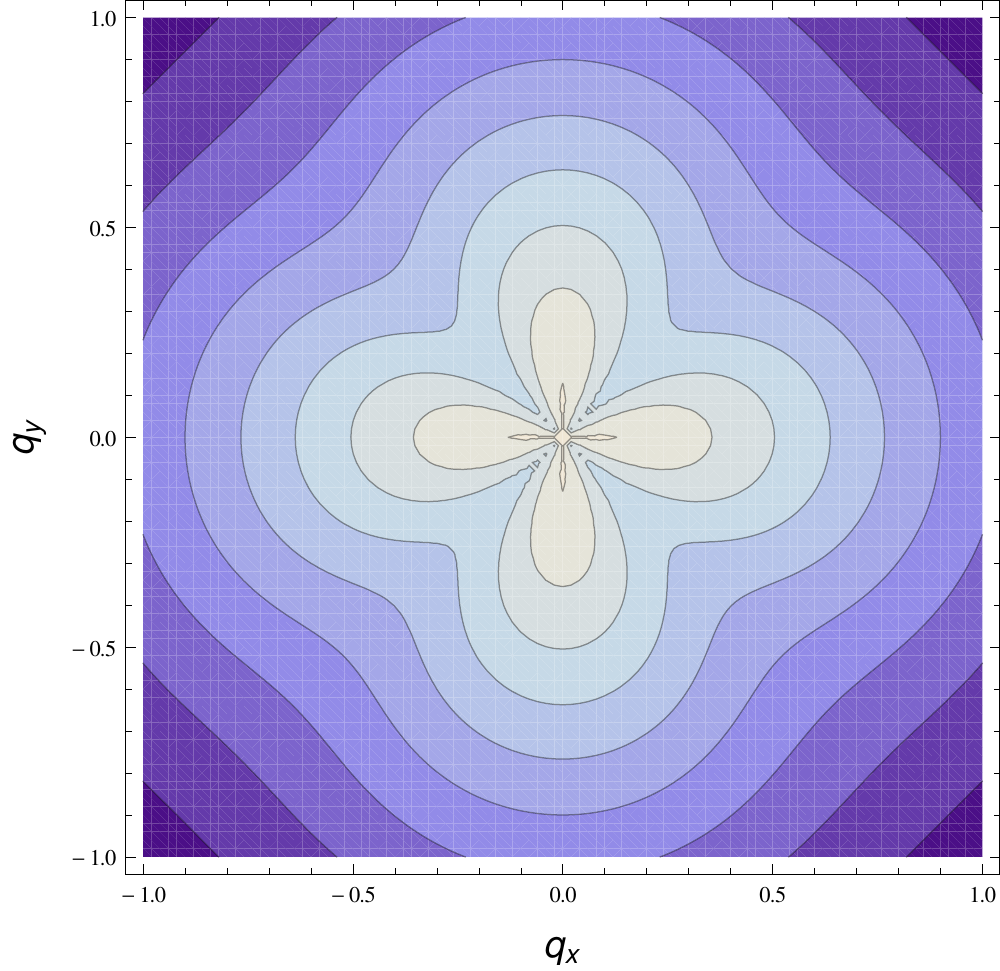}}
\caption{Density plot of the metamagnetic correlation function $\chi({\bf q})$ of Eq.~\eqref{MetamagCorr} in the limit $|{\bf q}| \to 0$ within the $(q_x,q_y) $ plane (a) in the absence, $\gamma_1 = 0$ and (b) in the presence, $\gamma_1 \neq 0$, of a magnetoelastic coupling for a certain choice of parameters ($r=2, \gamma_1^2/K_0 = 1.3,  \gamma_1^2/(C_{11} - C_{12}) = 0.7$, and $\gamma_1^2/C_{44} = 0.5$). For $\gamma_1 \neq 0$ the metamagnetic fluctuations inherit the spatial anisotropy of the phonon spectrum with a characteristic four-fold symmetry for the cubic crystal lattice.}
\label{fig:MetamagneticCorrelation}       
\end{figure}

Whereas the phonons remain non-critical at the isostructural QCEP they nevertheless renormalize the metamagnetic fluctuation spectrum. In order to simplify the following discussion, we again assume a cubic crystal where the metamagnetic fluctuations couple linearly to the trace of the strain tensor. The quadratic part of the theory \eqref{TheoryMetamagnetism} in the presence of an elastic coupling to the phonon modes $u_i$, see Eq.~\eqref{decomposition}, reads
\begin{align}
\mathcal{L}^{(2)} = \frac{1}{2} \phi \left[r - \nabla^2 \right] \phi - \gamma_1 \phi \partial_i u_i + \frac{1}{2} \partial_i u_j C_{ijkl} \partial_k u_l.
\end{align}
The static correlator of the metamagnetic fluctuations $\chi({\bf q}) = \langle \phi \phi\rangle_{\bf q}$ for $\gamma_1 = 0$ possesses a standard Ornstein-Zernicke form
\begin{align}
\chi_0({\bf q}) = \frac{1}{r + {\bf q}^2}
\end{align}
at $h=0$ and in lowest order in $u$. A finite linear magnetoelastic coupling $\gamma_1$ leads to a modification of the momentum dependence,
\begin{align}\label{MetamagCorr}
\chi({\bf q}) = \frac{\chi_0({\bf q})}{1 - \gamma_1^2 \chi_0({\bf q}) q_i (M^{-1})_{ij} q_j}
\end{align}
where $M^{-1}$ is the inverse of the dynamical matrix $M_{jl}({\bf q}) = q_iC_{ijkl} q_k$, see section \ref{subsec:basics}. For a cubic lattice the dynamical matrix reads explicitly
\begin{align}
M_{ij} = (C_{12} +C_{44}) q_i q_j + \Big(C_{44} {\bf q}^2+ (C_{11} - C_{12} - 2 C_{44}) q^2_i \Big) \delta_{ij}.
\end{align}
where  the last term should here not be summed over repeated indices. The bare modulus of the previous section is here identified with $K_0 = \frac{1}{3}(C_{11} + 2C_{12})$. Due to the phonon renormalization, the correlation function $\chi({\bf q})$ depends on the orientation of the momentum $\hat q$ even in the limit $|{\bf q}|  \to 0$. This is illustrated in Fig.~\ref{fig:MetamagneticCorrelation}. The inverse $1/\chi({\bf q})$ assumes for $|{\bf q}|  \to 0$ the limiting values
\begin{align}
\lim_{|{\bf q}|  \to 0}\chi^{-1}({\bf q}) = \left\{\begin{array}{lll}
r - \frac{\gamma_1^2}{K_0 + 2 (C_{11} - C_{12})/3} & \quad \mbox{for} & \hat q || \langle 100 \rangle\\
r - \frac{\gamma_1^2}{K_0 + \frac{1}{6} (C_{11} - C_{12}) +C_{44}} & \quad \mbox{for} & \hat q || \langle 110 \rangle\\
r - \frac{\gamma_1^2}{K_0 + \frac{4}{3} C_{44}} & \quad \mbox{for} & \hat q || \langle 111 \rangle.
\end{array}
 \right.
\end{align}
This explicit calculation also nicely demonstrates that the fluctuations indeed remain gapped at the isostructural QCEP that is realized at $r = r^* \equiv \gamma_1^2/K_0$. Due to the stability conditions $C_{11} - C_{12} > 0$ and $C_{44} > 0$, the gap $\lim_{|{\bf q}|  \to 0}\chi^{-1}({\bf q})$ remains finite for all directions of $\hat q$ even at the isostructural QCEP.

Such an asymmetry of the magnetic fluctuations is generically expected close to a metamagnetic end point and could, in principle, be detected with the help of neutron scattering. A similar asymmetry has been discussed and experimentally verified in the case of critical piezoelectric ferroelectrics \cite{Cowley76}.

\section{Critical Mott end point at finite temperature}
\label{sec:MEDP}

\begin{figure}
\centering
\resizebox{0.4\columnwidth}{!}{
 \includegraphics{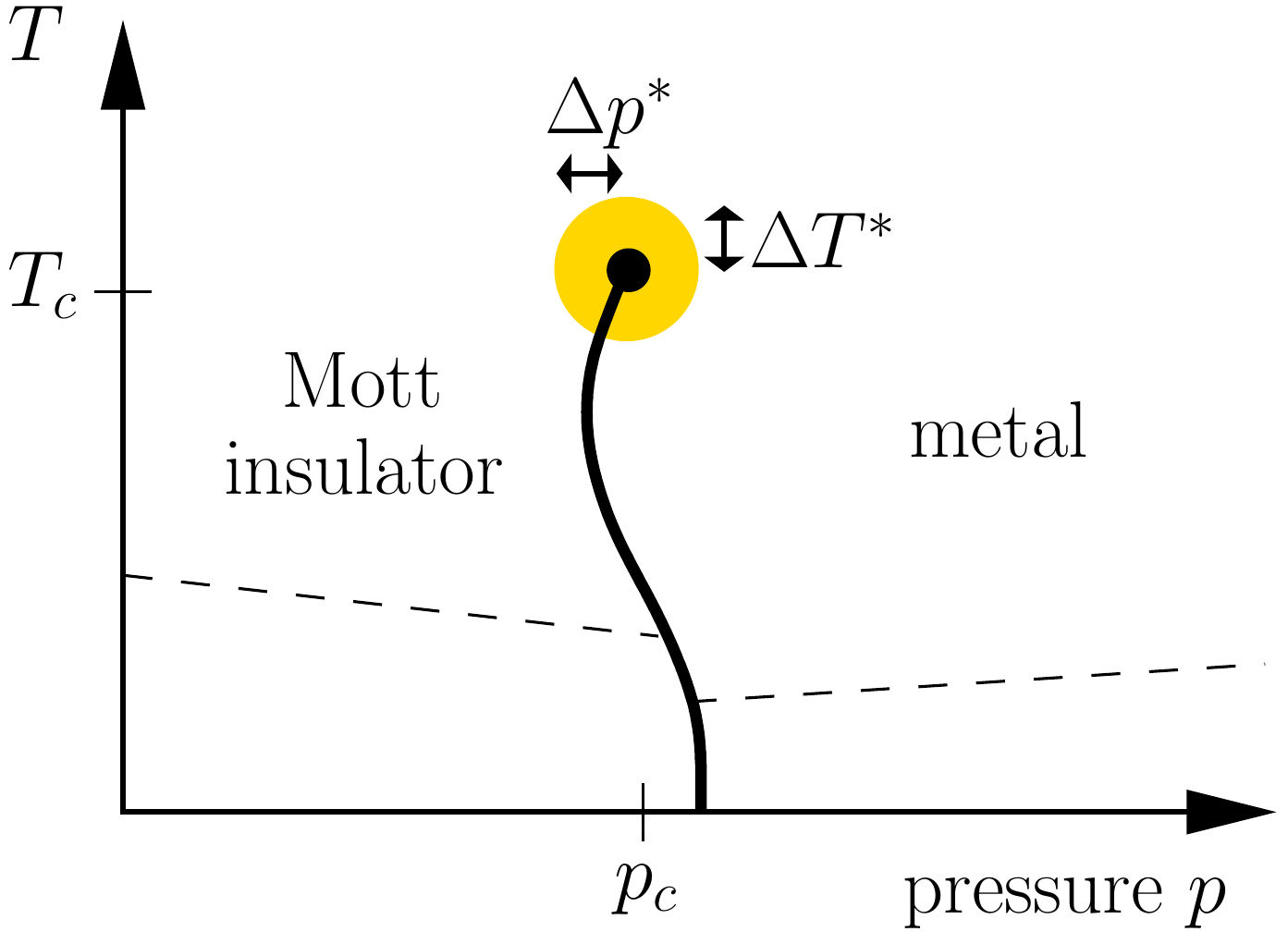} }
\caption{
Phase diagram of a system with a line of first-order Mott metal-insulator transitions that terminates in a second-order critical end point at $(p_c, T_c)$. Close to this end point elastic criticality prevails indicated by the yellow shaded regime. The criticality of the Mott end point therefore belongs to the universality class of a classical solid-solid end point characterized by mean-field Landau criticality \cite{Zacharias12}.
}
\label{fig:MottPD}       
\end{figure}

Strongly correlated materials with a conduction band close to half-filling might exhibit a Mott metal-insulator transition as a function of doping or applied pressure \cite{Imada98}. When all lattice sites are on average occupied by a single electron, their motion is suppressed by a strong Coulomb repulsion that energetically disfavours to have a site doubly occupied giving rise to a Mott insulating state.  The application of pressure, however, enhances the overlap integral of electron wavefunctions on adjacent lattice sites and thus the kinetic energy. For sufficiently large pressures the kinetic energy prevails resulting in a Mott metal-insulator transition at some critical pressure $p_c$.  

This transition is first-order at low temperatures resulting in a line of first-order transitions in the $(T,p)$ temperature-pressure plane. This line ends at a second-order critical end point at a finite $T_c$. For temperatures $T>T_c$ there remains only a crossover so that the system can be smoothly transformed from an insulator to a metal by varying pressure and temperature, see Fig.~\ref{fig:MottPD}. The topology of the phase diagram resembles the one of the liquid-gas transition and, correspondingly, it has been discussed theoretically and experimentally whether the Mott end point also belongs to the Ising universality class \cite{Castellani79,Kotliar00,Jayaraman:1970,Fournier03,Limelette03,Kagawa05,Kagawa09,Lefebvre00,Limelette03b,ToyotaBook,Papanikolaou08,deSouza07,Bartosch10,Imada05,Imada10,Sentef11,Semon12}.

However, we will argue that this is in fact not the case. As the Mott transition is susceptible to pressure tuning its conjugate, i.e., the volume will exhibit a jump across the line of first-order transitions. Correspondingly, the black line in the phase diagram of Fig.~\ref{fig:MottPD} does not only separate the Mott insulator from the metal but also two isostructural solids with different volumes. Consequently, its termination point is not in the liquid-gas, i.e., Ising universality but it is rather a classical critical solid-solid end point characterized by mean-field Landau criticality as  already pointed out in early work by Jayaraman {\it et al.} \cite{Jayaraman:1970}.

In this section, we will discuss the critical behavior close to the finite-temperature Mott end point in some detail, especially as in this case quantitative estimates of the importance of the lattice coupling are available. 
We will not consider the corresponding Mott quantum phase transition at $T=0$ here but, in contrast to the previous sections, strictly limit ourselves to the classical transition at finite $T_c>0$. We will take it for granted here that the finite $T$-end point of the line of first-order Mott metal-insulator transitions on an incompressible lattice is indeed described by the classical Ising model,
\begin{align} \label{IsingModel}
\mathcal{L} = \frac{1}{2} \phi (r-\nabla^2)\phi + \frac{u}{4!} \phi^4 - h \phi, 
\end{align}
with the Ising order parameter $\phi$ and pressure and temperature dependent tuning parameters $h = h(p,T)$ and $r = r(p,T)$. The end point of the incompressible lattice obtains for 
pressure and temperature values for which $h= r  =0$. 
The interaction $u$ gives rise to a Ginzburg scale  \cite{LandauBook5} that defines a crossover between Mott mean-field behavior away from the end point, that is described by Landau theory, and non-trivial critical behavior ascribed to the Ising universality class.   
The coupling of $\phi$ to the crystal lattice, Eqs.~\eqref{ElastoCoupling}, eventually gives rise to another crossover to a regime governed by critical elasticity. Depending on the relative strength of the elastic coupling, one can distinguish two scenarios that we will discuss in the following. 

\begin{figure}
\centering
(a) \resizebox{0.35\columnwidth}{!}{
 \includegraphics{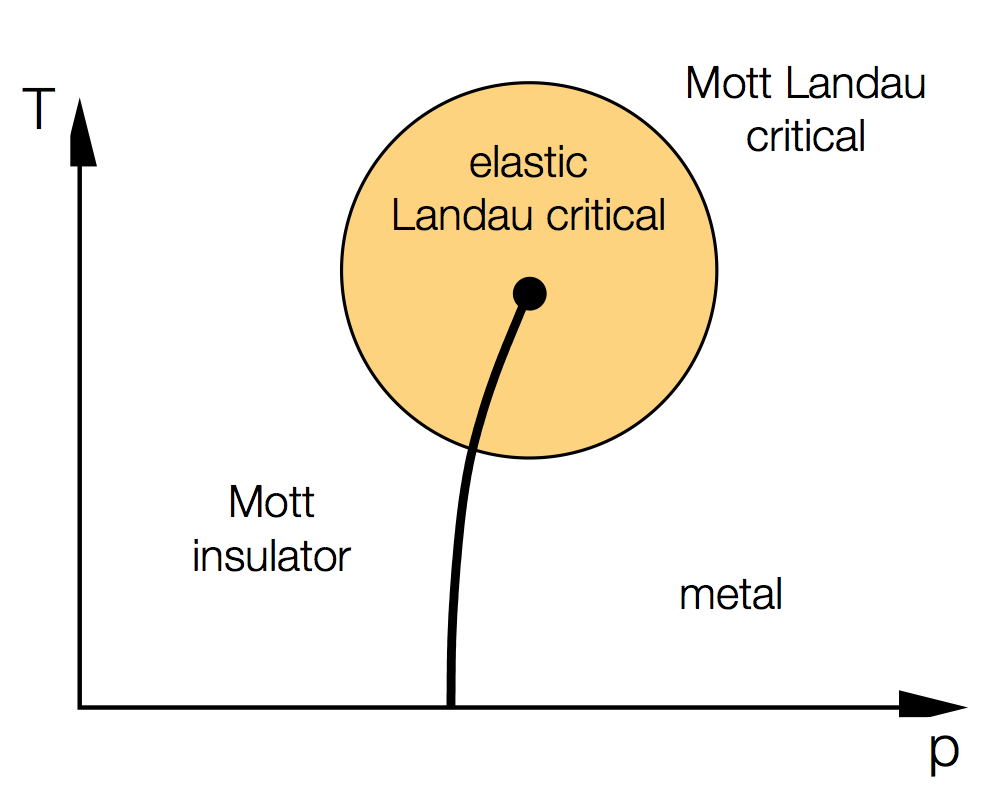}}
 \qquad\qquad (b)
 \resizebox{0.35\columnwidth}{!}{
 \includegraphics{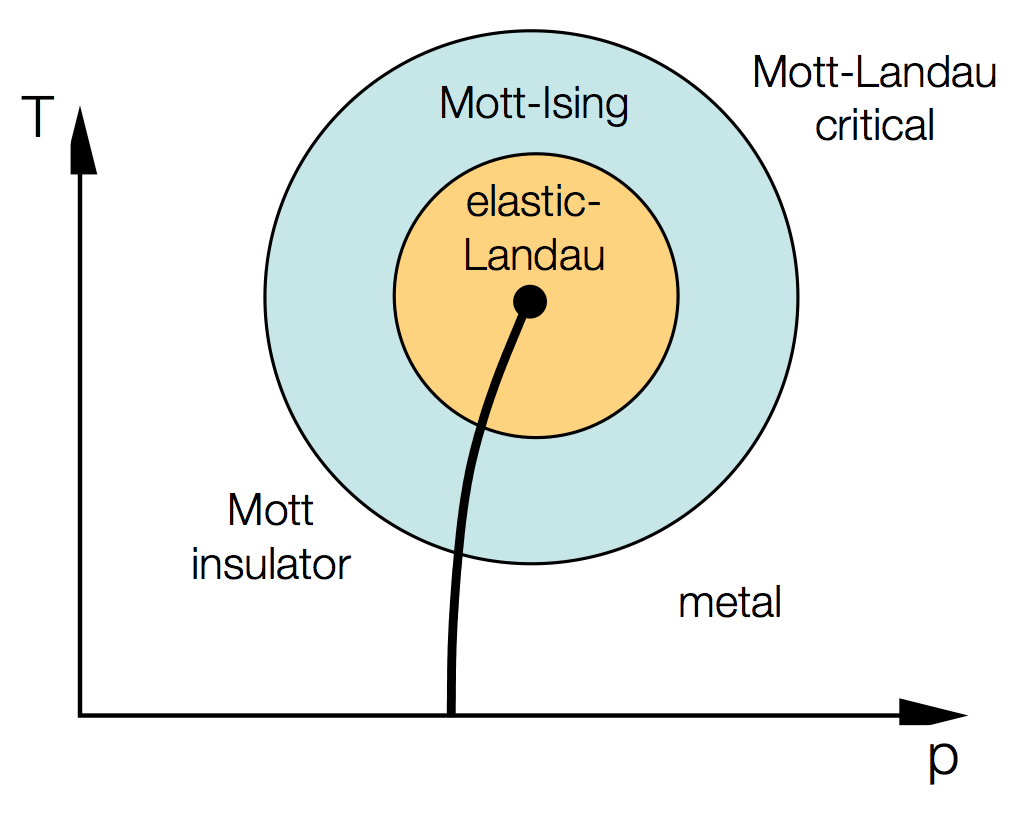}}
\caption{Sketch of two scenarios for crossovers upon approaching the Mott end point. (a) For relatively strong elastic coupling, there is a single crossover into a regime (yellow shaded) governed by elastic criticality. (b) For a relatively weak elastic coupling, first a crossover occurs into a regime (blue shaded) governed by the non-trivial Ising universality class, and afterwards a second crossover takes place into a regime (yellow shaded) controlled by elastic criticality. While scenario (a) 
applies to V$_2$O$_3$,  scenario (b)
is applicable to $\kappa$-(BEDT-TTF)$_2 X$. }
\label{fig:scenarios}       
\end{figure}

\subsection{Strong elastic coupling: Mott-Landau--to--elastic-Landau crossover}

For a relatively strong elastic coupling, there is just a single crossover from the mean-field Mott-Landau regime of the theory \eqref{IsingModel} to the regime in the immediate vicinity of the solid-solid end point that is governed by crystal elasticity, see Fig.~\ref{fig:scenarios}(a). This scenario can be captured within simple mean-field theory and is described by the potential,
\begin{align} \label{MottPotSc1}
\mathcal{V}(\phi,E) = \frac{r}{2} \phi^2 + \frac{u}{4!} \phi^4 - h \phi + \frac{1}{2} \gamma_2 E \phi^2 - \gamma_1 E \phi + \frac{K_0}{2} E^2 - (p-p_c) E.
\end{align}
The curvature tensor $\partial_\alpha \partial_\beta \mathcal{V}(\phi,E)$ with $\alpha,\beta = E, \phi$ at $h=0$ and $p=p_c$ 
for $\phi=0$ possesses a vanishing eigenvalue at $r^* = \gamma_1^2/K_0$. If the end point is approached for $p=p_c$ and $h = 0$ as a function of temperature, the Mott end point located at $r=0$ for the incompressible lattice is preempted by an isostructural solid-solid end point at $r = r^* > 0$. For the strong elastic coupling scenario, the scale $r^*$ in addition preempts the Ginzburg crossover present for the Mott transition on the incompressible lattice. As a result, the critical properties are always captured by Landau mean-field theory. 

Nevertheless, one could define a crossover between a Mott-Landau regime where the elastic coupling is still perturbative and a elastic Landau regime where Hooke's law breaks down. Solving the classical equations, $\partial_\phi \mathcal{V}(\phi,E) = 0$ and $\partial_E \mathcal{V}(\phi,E) = 0$, for the macroscopic strain, say, at criticality $r = r^*$, $h=0$, setting $\gamma_2 = 0$ for simplicity, one finds that for pressure values $|p-p_c| \gg \Delta p^*$ with $\Delta p^* = \gamma_1^2/\sqrt{K_0 u}$, the response is linear, $E \approx (p-p_c)/K_0$ but becomes non-linear, $E \sim |p-p_c|^{1/3}$, for $|p-p_c| \ll \Delta p^*$. The pressure scale $\Delta p^*$ defines the width of the Landau critical regime along the pressure axis that is governed by crystal elasticity (orange shaded regime in Fig.~\ref{fig:scenarios}(a)).

The scenario of a strong elastic coupling with a single crossover between a critical Mott-Landau to a elastic Landau mean-field regime might be applicable to V$_2$O$_3$. In Ref.~\cite{Limelette03} the critical behavior of Cr-doped V$_2$O$_3$, i.e., (V$_{0.989}$Cr$_{0.011}$)O$_3$, close to its end point was experimentally investigated with the help of conductivity measurements. Its behavior could be explained with the help of mean-field exponents in a large range of pressures and temperatures except in the immediate vicinity of the critical end point. It has been speculated \cite{Limelette03} that the deviations  observed very close to the critical pressure in a narrow range $\Delta p/p_c \approx 3 \permil$ 
indicate a Ginzburg crossover to non-trivial Ising exponents. However, detailed measurements of various sound velocities by Nichols {\it et al.} \cite{Nichols81} have revealed that the undoped compound V$_2$O$_3$, 
that translates to an effective distance from the critical end point of $\Delta p/p_c \approx 1.7 \%$ in pressure units, 
possesses already a substantial softening of the corresponding bulk modulus $K$. This modulus $K$ denoted as $\lambda_4$ in Ref.~\cite{Nichols81} has a deep minimum around 540 K 
where it drops to around $26 \%$ of its value at room temperature. From our  considerations above follows the expectation that the value of this $\lambda_4$ modulus indeed vanishes exactly at the critical end point. The fact that it is so strongly reduced 
 in the pure compound indicates that V$_2$O$_3$ already accesses the elastic critical regime  (yellow shaded in Fig.~\ref{fig:scenarios}(a)) as a function of temperature. In turn, it follows that most of the experiment of Ref.~\cite{Limelette03} where the conductivity at the critical temperature was tuned with the help of pressure is located in this elastic critical regime. This suggests that the end point in this material is always governed by Landau mean-field theory and, in particular, that the deviations mentioned above are not caused by a Ginzburg crossover but are rather attributable to some other origin as, for example, the uncertainty in determining the precise form of scaling variables like $r, h$ or $p-p_c$ in Eq.~\eqref{MottPotSc1}.

\subsection{Weak elastic coupling: Mott-Ising--to--elastic-Landau crossover}
\label{subsec:WEC}

\begin{figure}
\centering
(a) \resizebox{0.35\columnwidth}{!}{
 \includegraphics{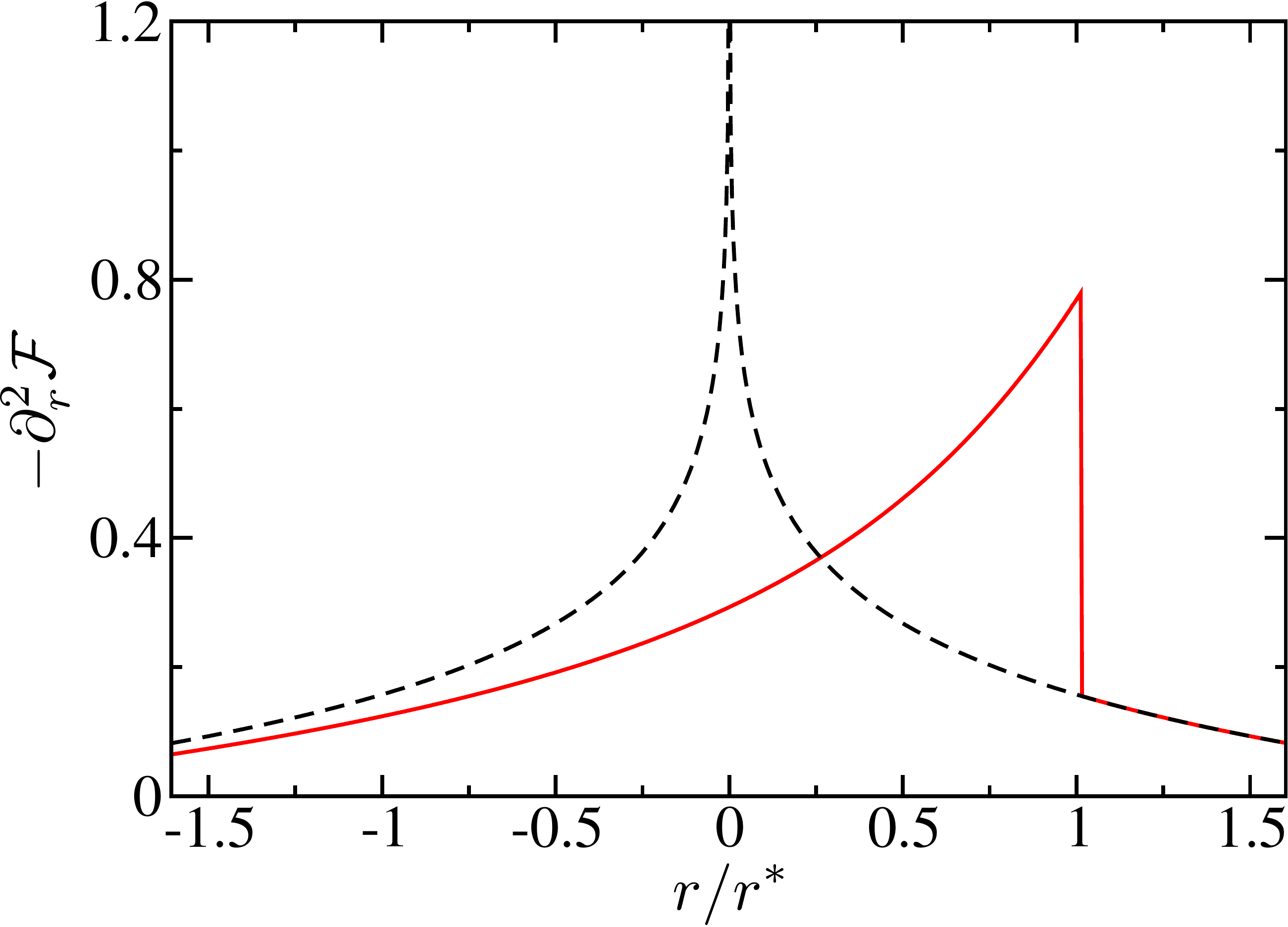}}
 \qquad\qquad (b)
 \resizebox{0.35\columnwidth}{!}{
 \includegraphics{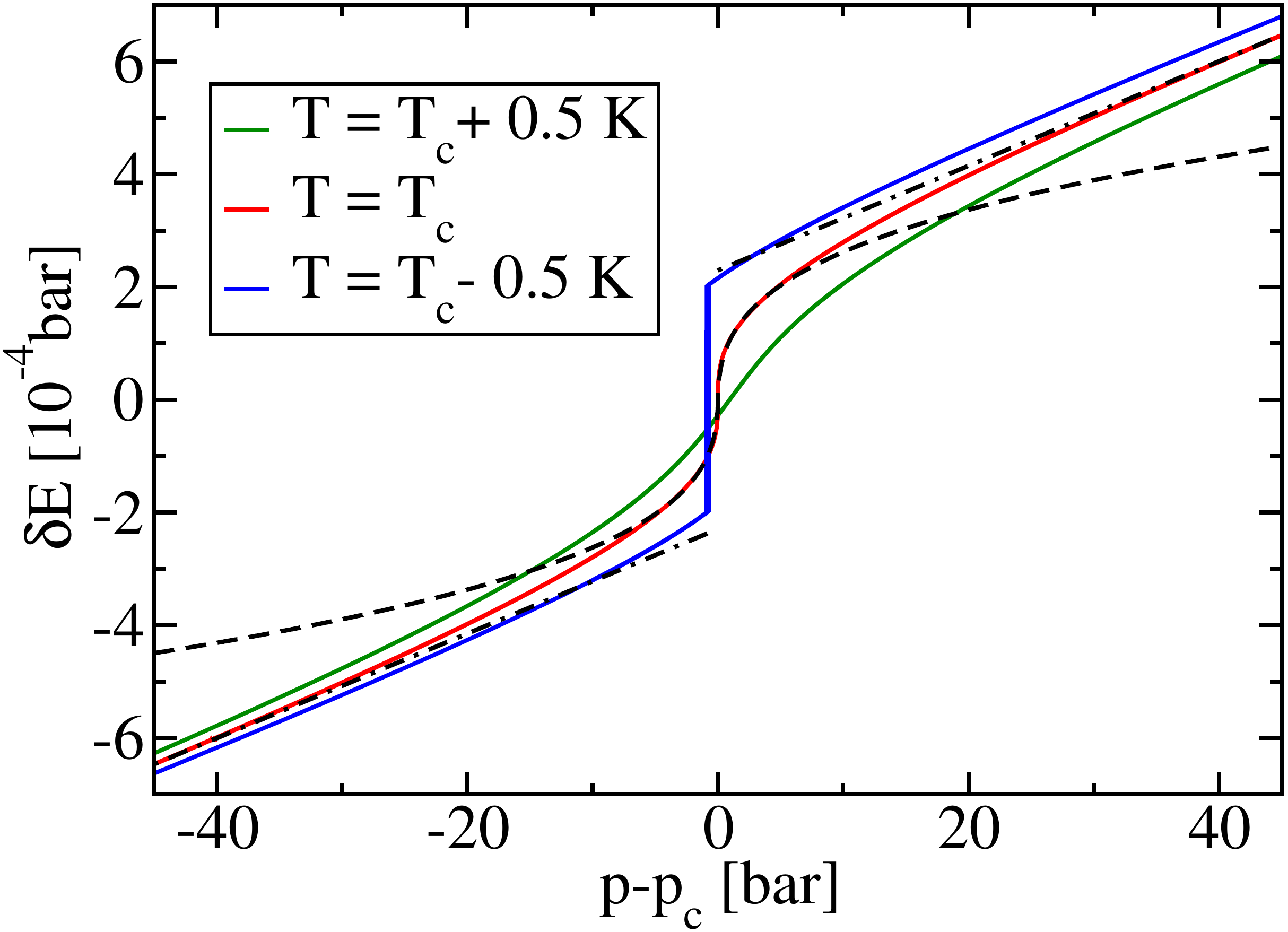}}
\caption{
(a) The second derivative $-\partial^2_{r} \mathcal{F}$ in dimensionless units at the critical pressure, $p_c$. It shows a pure mean-field jump (solid line) as a function of $r$ preempting the logarithmic Ising singularity (dashed line). (b) Macroscopic strain as a function of $p-p_c$ for various temperatures $T$, see text.}
\label{fig:MottResults}       
\end{figure}

For a relatively weak elastic coupling, on the other hand, there are two crossovers, see Fig.~\ref{fig:scenarios}(b), upon approaching the critical end point. First, a crossover at the Ginzburg scale occurs into a regime where the critical behavior is governed by non-trivial Ising exponents (blue shaded regime) until a second crossover occurs into the elastic critical regime (yellow shaded regime) described by Landau theory. 

In this section, we concentrate on the latter, Mott-Ising to elastic Landau crossover from the blue to yellow shaded regime in Fig.~\ref{fig:scenarios}(b). For its description, we can adopt the approach of section \eqref{sec:solid-solidMetamag}, neglect the phonons and integrate out the Mott-Ising degrees of freedom. This yields an effective potential for the macroscopic strain singlet 
\begin{align} \label{PotMottSce2}
\mathcal{V}(E) = \frac{K_0}{2} E^2 + \sigma_0 E + f_{\rm sing}(r, h + \gamma_1 E)
\end{align}
where we included already the corresponding singlet stress $\sigma_0$. The function $f_{\rm sing}$ is here governed by the Ising fixed-point and not known in closed form for general spatial dimension $d$. 

An extended Mott-Ising regime seems to be realized close to the Mott end point of the quasi two-dimensional charge-transfer salt $\kappa$-(BEDT-TTF)$_2 X$. Extensive investigations have demonstrated the critical properties in the vicinity of its end point is controlled by non-trivial critical behavior probably of Ising type \cite{Kagawa05,Kagawa09,Lefebvre00,Limelette03b,ToyotaBook,Papanikolaou08,deSouza07,Bartosch10}.
It is suggestive that the reduced spatial dimensionality of the fluctuation in this material promotes the Ginzburg crossover and thus the Mott-Ising regime before the elastic crossover sets in.

The available data of thermal expansion \cite{deSouza07,Bartosch10} on $\kappa$-(BEDT-TTF)$_2 X$ allowed us in Ref.~\cite{Zacharias12} to determine the parameters of the potential \eqref{PotMottSce2} and to theoretically describe the expected second crossover to elastic criticality in this material. 
Motivated by its quasi-two-dimensional structure, we followed Bartosch {\it et al.} \cite{Bartosch10} and used in Eq.~\eqref{PotMottSce2} the free energy, $f_{\rm sing}$, of the two-dimensional Ising model, $d=2$, in the asymptotic scaling regime, that can be computed numerically using the results of Ref.~\cite{Fonseca2003}. Minimizing \eqref{PotMottSce2} with respect to $E$ then yields the free energy $\mathcal{F}$.

Results from Ref.~\cite{Zacharias12} for the $T$-dependence of the second derivative $-\partial^2_{r} \mathcal{F}$ at the critical pressure $p_c$, which will govern the specific heat, are shown in Fig.~\ref{fig:MottResults}(a). Without the elastic coupling, $\gamma_1=0$, this derivative exhibits the logarithmic singularity of the 2d Ising model (dotted line), that is however cutoff for finite $\gamma_1 > 0$ (solid line) and replaced by a pure mean-field jump \cite{Levanyuk:1970}.
The expected pressure dependence of the strain $E$ for various temperatures is shown in Fig.~\ref{fig:MottResults}(b).  Whereas for $T < T_c$ the change of strain $\delta E$ as a function of $p-p_c$ exhibits a first-order jump, it only exhibits a crossover for $T > T_c$. For $T=T_c$, finally, it behaves as $\delta E \sim |p-p_c|^{1/3}$ indicating a breakdown of Hooke's law. 

Our quantitative analysis allowed us, in particular, to estimate the width of the elastic critical regime in $\kappa$-(BEDT-TTF)$_2 X$ along the pressure and temperature axis, $\Delta p^*$ and $\Delta T^*$, respectively, 
\begin{align}
\Delta p^* = 45\, {\rm bar},\quad \Delta T^* = 2.5\, {\rm K}.
\end{align}
These values suggest that it should be experimentally feasible to detect the crossover from the Mott-Ising to the elastic critical regime, and thus investigate the change in universality class upon approaching the solid-solid end point of the Mott transition.
\section{Conclusions}

The characteristic feature of crystals is their shear rigidity which distinguishes solids in particular from liquids or gases.
This shear rigidity and the corresponding long-range shear forces can in certain cases fundamentally alter the critical behavior close to a second-order phase transition. An especially drastic example is the elastic instability associated with a divergence of the bulk compressibility close to a solid-solid critical end point discussed in section \ref{subsec:IsostructuralElTr}. While the bulk compressibility diverges the phonons remain non-critical because the shear stiffness ensures that the phonon velocities all remain finite. As a result, the critical behavior of solid-solid end points do not possess a diverging correlation length and are described by true mean-field behavior. This is to be contrasted with the properties of the liquid-gas end point where the diverging compressibility implies the vanishing of the sound velocity, that in turn leads to a diverging correlation length and to non-trivial critical behavior belonging to the Ising universality class. 

In this paper we have discussed cases where the shear rigidity of solids strongly affects quantum phase transitions.
Besides genuine elastic transitions, this physics is of importance for phase transitions that are primarily driven by magnetic, ferroelectric or electronic degrees of freedom. A first such case are critical end points of "liquid-gas type" described by an Ising order parameter. 
In such a case, the coupling of the order parameter to the strain field of the crystal will change the universality class and critical elasticity prevails close to the critical point. We discussed two examples, the metamagnetic quantum critical end point in section \ref{sec:MQCEP} and the Mott end point at finite temperature in section \ref{sec:MEDP}, both of which are governed by the properties of solid-solid end points. 

Elastic quantum phase transitions associated with a change of crystal symmetry were discussed in section \ref{subsec:SSBET}. In case of spontaneous crystal symmetry-breaking, phonon velocities in certain crystallographic directions vanish resulting in critical thermodynamics that is however still governed by a Gaussian fixed-point. We discussed the corresponding quantum critical signatures and the resulting divergence of the Gr\"uneisen parameter. This quantum critical elasticity might be relevant, for example, for the pnictide superconductors at its continuous tetragonal-to-orthorhombic transition close to optimal doping \cite{Zacharias2014}.

Studies of the thermal expansion, compressibility and of ultrasound are especially useful to address the physics discussed in this review. With the help of such measurements, the strength of the elastic coupling can be determined and the crossover to the regime governed by critical elasticity can be quantitatively estimated, for a specific example see section \ref{subsec:WEC}. Furthermore, they allow to identify directly a breakdown of Hooke's law or the vanishing of sound velocities that are characteristic features of elastic phase transition.  
\\[3em]
{\small This is a review on work (Refs.~\cite{Zacharias2014,Zacharias12,ZachariasPhD,Zacharias13,Weickert10}) that was supported by the DFG through FOR 960. We acknowledge a collaboration with I. Paul and L. Bartosch on topics presented here. }

\end{document}